\documentclass[12pt, a4paper]{article}
\usepackage{graphicx}
\graphicspath{ {./} }
\usepackage{amsmath}
\usepackage{amsfonts}
\usepackage{amssymb}
\usepackage{geometry}
\usepackage{longtable}
\usepackage{array}
\usepackage{enumitem}
\usepackage{microtype}
\usepackage{hyperref}
\usepackage{adjustbox}

\usepackage{tikz}
\usetikzlibrary{
    positioning,
    shapes.geometric,
    arrows.meta,
    backgrounds,
    fit
}

\definecolor{userquerycolor}{HTML}{e1f5fe}
\definecolor{candidatesetcolor}{HTML}{c8e6c9}
\definecolor{fusioncolor}{HTML}{fff3e0}
\definecolor{embeddingcolor}{HTML}{fce4ec}

\tikzstyle{process} = [rectangle, draw, fill=blue!10, text width=3.2cm, minimum height=1cm, align=center]
\tikzstyle{decision} = [diamond, draw, fill=green!20, text width=2.5cm, minimum height=1cm, align=center, aspect=1.5]
\tikzstyle{database} = [cylinder, draw, fill=orange!20, shape aspect=0.5, minimum height=1cm, text width=3cm, align=center]
\tikzstyle{result} = [rectangle, rounded corners, draw, fill=purple!10, text width=3.2cm, minimum height=1cm, align=center]
\tikzstyle{subgraph_box} = [rectangle, rounded corners, draw=gray, fill=gray!5, inner sep=0.5cm]
\tikzstyle{arrow} = [thick, ->, >=Stealth]
\tikzstyle{userQueryStyle} = [process, fill=userquerycolor, rounded corners]
\tikzstyle{candidateSetStyle} = [result, fill=candidatesetcolor]
\tikzstyle{fusionStyle} = [process, fill=fusioncolor, shape=star, star points=6, star point ratio=0.8]
\tikzstyle{embeddingStyle} = [process, fill=embeddingcolor]

\hypersetup{
    colorlinks=true,
    linkcolor=blue,
    filecolor=magenta,
    urlcolor=cyan,
    pdftitle={Irec: A Metacognitive Scaffolding for Self-Regulated Learning through Just-in-Time Insight Recall},
    pdfauthor={Xuefei Hou and Xizhao Tan},
}

\title{Irec: A Metacognitive Scaffolding for Self-Regulated Learning through Just-in-Time Insight Recall: A Conceptual Framework and System Prototype}
\author{Xuefei Hou \and Xizhao Tan}
\date{\today}

\begin{document}

\maketitle

\begin{abstract}
In an increasingly complex web of knowledge, the core challenge for learners has shifted from knowledge acquisition to effective Self-Regulated Learning (SRL)—how to plan, monitor, and reflect on one's own learning process. However, existing digital learning tools show significant deficiencies in supporting metacognitive reflection, a key component of SRL. Traditional Spaced Repetition Systems (SRS) combat forgetting through de-contextualized review, yet they overlook the fundamental role of context in knowledge retrieval. Meanwhile, Personal Knowledge Management (PKM) tools, while supporting networked thinking, have their seamless integration into the learning workflow hindered by high manual maintenance costs.

To address these challenges, this paper introduces a novel learning support paradigm: "Insight Recall." This paradigm conceptualizes the "retrieval of personal past insights triggered by context" as a metacognitive scaffold aimed at promoting SRL. We further formalize the "Insight Recall" paradigm using the Just-in-Time Adaptive Intervention (JITAI) framework, providing a rigorous theoretical model for its design and analysis. Based on this, we design and implement a prototype system named \texttt{Irec} to operationalize the paradigm and demonstrate its technical feasibility. At the core of \texttt{Irec} is a dynamic knowledge graph driven by the user's personal learning history. When a user encounters a new problem (a JITAI decision point), the system employs a hybrid retrieval engine—dynamically adjustable according to the user's chosen learning mode (a JITAI decision rule)—to accurately recall the most relevant personal "insights" from the graph. Subsequently, the system utilizes a large language model to perform a deep similarity assessment and filter the recalled insights, ultimately presenting the highly relevant metacognitive scaffold in a just-in-time manner. To automate this process and reduce the user's cognitive load, \texttt{Irec} integrates a human-in-the-loop knowledge graph construction pipeline based on a Large Language Model (LLM). To further overcome the risks of cognitive fixation and error reinforcement that may arise from a single reflection partner, we propose and integrate an optional "Guided Inquiry" module. Users can, with a single click, use the current problem and recalled insights as context to engage in a Socratic dialogue with a domain-expert LLM, thereby deepening a simple "review" into an "expert-guided inquiry." Finally, we use a detailed illustrative scenario to intuitively demonstrate the core workflow of \texttt{Irec}. The contribution of this paper is to provide a solid theoretical framework and a usable system platform for designing the next generation of intelligent learning systems aimed at enhancing learners' metacognition and self-regulation capabilities.

\vspace{1em}
\noindent\textit{\textbf{A Note on This Version (v1):} This manuscript presents the core conceptual framework and system architecture. A subsequent version (v2) is planned to include finalized system flowcharts, a public GitHub repository with the source code, and a reproducibility package detailing the prompts, models, and testing guidelines used.}

\vspace{1em}
\end{abstract}

\section{Introduction}

In an era where information is readily available, the bottleneck of learning is no longer the acquisition of information, but its effective internalization and integration. Learners, especially those engaged in fields with highly interconnected knowledge points and strict logical chains such as mathematics, programming, and law, face a common dilemma: how to weave isolated pieces of knowledge into a coherent and flexible cognitive network. The core of this process is \textbf{Self-Regulated Learning (SRL)}—a cyclical process in which learners actively participate in planning, monitoring, controlling, and reflecting on their own cognition, motivation, and behavior. Among the many components of SRL, \textbf{metacognitive reflection} (i.e., "thinking about one's own thinking process") is considered key to achieving deep understanding and transfer of knowledge.

However, existing digital learning tools generally fall short in providing effective support for this high-order cognitive activity. On one hand, memory tools represented by Spaced Repetition Systems (SRS) are philosophically rooted in Ebbinghaus's forgetting curve. They have achieved great success in reinforcing the memory of atomized facts (such as words and formulas) by algorithmically scheduling review times. But their core mechanism is \textbf{de-contextualized}. Knowledge points are stripped from their original problem-solving context and presented in isolation at predetermined times. This contradicts one of the most fundamental findings in cognitive psychology—the \textbf{Encoding Specificity Principle}. This principle states that the effectiveness of memory retrieval largely depends on the degree of match between the cues at retrieval and the context at encoding. A review card about an infinite series that suddenly pops up while solving a calculus problem may temporarily activate the memory, but it interrupts the coherent flow of thought and fails to promote deep connections between new and old knowledge.

On the other hand, modern Personal Knowledge Management (PKM) tools like Obsidian and Logseq have greatly empowered networked thinking by introducing bidirectional links, allowing users to build personal knowledge graphs. This is philosophically aligned with the goal of constructing a cognitive network. However, the great flexibility of these tools also brings their core challenge in the learning domain: \textbf{high manual maintenance costs}. Users need to act like librarians, designing and maintaining complex tagging systems and link structures themselves. This cognitive load diverts learners' energy from the learning content itself to managing the learning tool, hindering the scalability and seamlessness of knowledge management.

More importantly, both types of tools fail to effectively support a cognitive activity that is crucial in real-world learning. Students often spend a lot of time on "problem-solving drills," yet what they often forget is not the answers to the problems, but the valuable "experiences" or "insights" gained while solving them. When this key experience is forgotten, they can only re-consolidate the knowledge by repeatedly doing more problems of the same type, leading to a great waste of learning efficiency. An ideal learning support system should be able to, when a learner faces a new challenge, precisely and appropriately reunite them with their past "aha moments," replacing inefficient "repetitive practice" with efficient "experience reuse."

To systematically address the above challenges, we introduce "Insight Recall," a learning paradigm designed to promote metacognitive reflection by presenting relevant past personal insights in a just-in-time manner. We argue that this personal historical review, triggered by a new context and centered on comparison and reflection, is a highly effective \textbf{Metacognitive Scaffold} that can effectively promote key aspects of self-regulated learning.

To put this paradigm into practice, we designed and implemented a prototype system called \texttt{Irec}. \texttt{Irec} aims to automate the capture, organization, and retrieval of learners' "insights," shifting the process of learning reinforcement from "user-actively-maintains, system-passively-schedules" to "system-actively-captures, context-triggers-recall." This significantly reduces the user's cognitive burden and creates the conditions for efficient metacognitive reflection.

The main contributions of this paper are conceptual and technical, specifically including:
\begin{itemize}
    \item \textbf{Conceptual Contribution:} Introduced and formalized the "Insight Recall" paradigm. This paradigm conceptualizes the "retrieval of personal past insights triggered by context" as a \textbf{metacognitive scaffold} aimed at promoting key processes of \textbf{Self-Regulated Learning}.
    \item \textbf{Theoretical Contribution:} Applied theoretical tools such as Activity Theory, Distributed Cognition, and the Just-in-Time Adaptive Intervention (JITAI) framework to provide a solid and multidimensional theoretical interpretation and formal modeling for the "Insight Recall" paradigm.
    \item \textbf{System Contribution:} Designed and implemented a prototype system \texttt{Irec} that supports "Insight Recall," featuring a three-path hybrid retrieval engine, demonstrating how to operationalize the "Insight Recall" paradigm and its theoretical model.
    \item \textbf{Illustrative Contribution:} Through a detailed illustrative scenario, intuitively demonstrated the core workflow and potential of \texttt{Irec}.
\end{itemize}

This paper aims to lay the theoretical and technical foundation for this paradigm and to provide a usable platform for future empirical research, rather than providing a final validation of the paradigm's effectiveness.

\section{Related Work}

The proposal of the "Insight Recall" paradigm aims to carve out a new niche at the intersection of existing digital learning tools. To clearly position the theoretical and systemic contributions of \texttt{Irec}, this chapter will conduct an in-depth and comprehensive analysis of related research fields from three core theoretical themes—contextualized learning, knowledge management automation, and the role of AI in education—thereby systematically elucidating the uniqueness and advancement of "Insight Recall."

\subsection{From De-contextualized Repetition to Context-Triggered Retrieval}

The theoretical cornerstone of Spaced Repetition Systems (SRS) is the Spacing Effect, which posits that learning distributed over time leads to more durable memory than massed learning. Tools like Anki and SuperMemo, using algorithms such as SM-2 to optimize review intervals, have achieved recognized success in strengthening long-term memory for "declarative knowledge" (i.e., knowledge of "what"). However, when applied to domains requiring deep understanding and procedural knowledge, the limitations of their core mechanism—\textbf{de-contextualization}—become apparent.

In an SRS, a knowledge point (e.g., the application of a mathematical theorem) is stripped from its original learning context (e.g., the process of solving a specific problem) and made into an atomized "question-answer" card. The timing of the review is determined by a temporal algorithm unrelated to the learner's current cognitive task. This model directly violates a fundamental principle in cognitive psychology, the \textbf{Encoding Specificity Principle}, which holds that the success of memory retrieval is highly dependent on the match between the retrieval context and the encoding context. When the system presents an isolated knowledge point at a moment unrelated to the current task, it may temporarily refresh the memory trace, but it fails to promote meaningful connections between that knowledge point and the new information the learner is currently processing. Broader research also confirms that context, whether it be the physical environment, emotional state, or the cognitive task itself, is an indispensable part of learning and memory.

This philosophy coincides with the rise of "Retrieval-Augmented Generation" (RAG) in contemporary artificial intelligence research. RAG systems significantly improve the accuracy and timeliness of large language models by retrieving relevant information from large knowledge bases before generating an answer. However, most RAG applications focus on retrieving factual knowledge from public, massive corpora (like Wikipedia). \texttt{Irec}'s uniqueness lies in applying the idea of RAG to a highly personalized, dynamically evolving personal knowledge graph. It retrieves not universal "facts" but the user's past "insights," thereby transforming the act of retrieval itself into a catalyst for metacognitive reflection.

Furthermore, the typical interaction pattern of SRS (see question - recall answer - compare with standard answer) primarily trains a passive "recognition-based" cognitive activity. However, learning that truly promotes deep understanding is often \textbf{Generative Learning}. Generative learning theory posits that when learners actively reorganize, integrate, summarize, and explain information, they construct a deeper and more lasting understanding. \texttt{Irec}'s "Insight Recall" is designed precisely to promote this generative activity. When the system presents a new problem context alongside a relevant historical "insight," it is not asking the user to passively recall a standard answer, but rather creating a cognitive situation that requires active thinking. The learner is naturally guided to perform high-order cognitive activities such as \textbf{Comparing}, \textbf{Classifying}, and \textbf{Self-explaining}, thus achieving a cognitive leap from "knowing what" to "knowing why and how." Therefore, the research focus should shift from "how to optimize context-free repetition intervals" to "how to design systems that can trigger meaningful, contextualized, and generative retrieval."

\subsection{From Manual Card Creation to Human-AI Collaborative Knowledge Co-construction}

Second-generation Personal Knowledge Management (PKM) tools, represented by Obsidian and Logseq, have made it possible to build personal knowledge networks by introducing bidirectional links and graph views. This is philosophically highly consistent with \texttt{Irec}'s concept of networked knowledge. These tools give users a great deal of freedom to capture, organize, and connect information. However, the other side of this freedom is a huge cognitive load.

In these general-purpose PKM tools, the quality of the knowledge network depends entirely on the user's own organizational skills and discipline. The user needs to act like a knowledge "artisan," spending a great deal of time and effort designing tag systems, creating links, and maintaining structural consistency. As the size and complexity of the knowledge base increase, the maintenance cost grows exponentially, easily leading to "tag chaos" or "link islands," ultimately making the knowledge graph difficult to navigate and utilize. This "manual card creation" bottleneck is the core obstacle preventing PKM tools from realizing their full potential in fast-paced, high-intensity learning scenarios.

\texttt{Irec}'s response is to introduce the technology of \textbf{Automatic Knowledge Graph Construction} into the personal learning domain. This is a fairly mature research direction in information retrieval and artificial intelligence, aiming to automatically extract entities and relations from unstructured or semi-structured data sources (like text) and build them into a structured knowledge graph. The "Insight Capture and Nodification" and "Intelligent Tag Mapping" modules in the \texttt{Irec} system are specific applications of this technical idea in a personal learning context.

Importantly, \texttt{Irec} does not pursue full automation. Given that current Large Language Models (LLMs) may still have inaccuracies when processing complex semantics, \texttt{Irec} incorporates a critical \textbf{Human-in-the-loop Validation and Refinement Layer} in its architecture. The LLM is responsible for "guiding" the extraction of structured "insights" and tag suggestions from unstructured learning notes, but the final confirmation, modification, and veto power always remains with the user. This design positions the system as a \textbf{human-AI collaborative knowledge co-construction partner}, where the role of AI is to lower the barrier to structuring and provide suggestions, while the role of the human is to make the final determination of meaning. This not only ensures the accuracy and personalization of the knowledge graph but also respects the learner's cognitive agency.

\subsection{From Instructive Tutoring to a Reflective Partner}

The application of Artificial Intelligence in Education (AIED) has traditionally been represented by \textbf{Intelligent Tutoring Systems (ITS)}. ITS usually play the role of an omniscient "tutor," with built-in expert knowledge models and student models, recommending optimal learning paths or practice problems by diagnosing students' knowledge weaknesses. The goal of an ITS is to guide the student towards a predetermined "correct" answer.

However, the role of AI in education is far more than this. Researchers have explored various different AI education paradigms. For example, \textbf{Learning Companions} shape AI as a "peer" who learns alongside the learner, stimulating learning motivation through social interaction. \textbf{Teachable Agents (TAs)}, on the other hand, design AI as a virtual student that needs to be "taught" by the user. In the process of teaching the TA, the learner is forced to clearly articulate and organize their own knowledge, thereby deepening their understanding through "learning by teaching," a process known as the "protégé effect."

\texttt{Irec} occupies a unique niche in this rich spectrum of AIED roles. It is neither a tutor, nor a peer or a student. Its core role is that of an efficient \textbf{"cognitive reflection support tool"}, or a perfect personal \textbf{"second brain"}. It does not presuppose any expert knowledge; its knowledge graph is entirely generated by the user's own learning trajectory. It does not provide answers to external problems; the "insights" it recalls are entirely the user's own past thoughts. The core intervention method of \texttt{Irec} is to create a \textbf{self-dialogue across time}: letting the you of today meet the thoughts of the you of yesterday.

The essence of this intervention is a \textbf{Metacognitive Scaffold}. Metacognitive scaffolds are external supports designed to guide learners in planning, monitoring, and evaluating their own learning process. When \texttt{Irec} presents a relevant historical insight, it is not providing an answer, but rather posing an implicit metacognitive question: "Look at how you thought about this problem in the past. What are the similarities and differences with your current thinking? What patterns can you abstract from this?". This support directly acts on the learner's self-regulated learning (SRL) cycle, particularly its monitoring and reflection phases. In recent years, research on AI-driven metacognitive scaffolds has been increasing, with numerous studies showing that using AI tools to provide personalized feedback and metacognitive support can significantly promote students' SRL abilities.

In summary, the "Insight Recall" paradigm carves a new path in the existing technological landscape by uniquely combining three elements: context-triggered retrieval, human-AI collaborative knowledge graph construction, and the role of AI as a metacognitive scaffold. It aims to build an intelligent learning environment that is more aligned with human cognitive principles and more conducive to promoting deep learning and self-regulation.

\section{Theoretical Framework and Model}

To transform the abstract concept of "Insight Recall" into a designable, analyzable, and implementable system framework, we first provide precise definitions of its core concepts. We then introduce Activity Theory and Distributed Cognition as theoretical lenses for interpretation, and finally, we use the Just-in-Time Adaptive Intervention (JITAI) model to formalize it.

\subsection{Core Concept Definitions}

In this framework, we define three core concepts. \textbf{Insight} specifically refers to the summary of problem-solving ideas, mental models, key techniques, or "aha moments" that a learner obtains when solving a specific problem and subjectively considers to be central or critical. It is a highly personalized cognitive product that carries traces of thought, rather than a standardized answer. In the \texttt{Irec} system, it is stored structurally in "ProblemCard" nodes of the knowledge graph. \textbf{Recall} does not refer to the passive memory retrieval of the human brain, but to an active, context-triggered presentation behavior of the system. When a new learning context (i.e., "context") appears, the system accurately retrieves the most historically relevant "insight" from the knowledge base and "recalls" it to the user's current cognitive field of view. In summary, \textbf{Insight Recall} is defined as an automated, context-aware learning support paradigm based on a personal knowledge graph and driven by artificial intelligence. Its core hypothesis is that the optimal time to consolidate and internalize past knowledge is not determined by a preset time algorithm, but when the learner encounters a new situation in the current task that is highly relevant to that knowledge. This review of personal past "insights," triggered by a new context, aims to stimulate higher-order metacognitive activities by promoting comparison, reflection, and abstraction, thereby achieving the networked construction and deep transfer of knowledge.

It is worth emphasizing that "Insight Recall" is fundamentally different from the "similar problem push" feature common in question banks. The goal of the latter is to provide consolidation practice with similar types of problems, while the goal of "Insight Recall" is to trigger heuristic reflection by reviewing personal past "aha moments." The former provides standardized problems, while the latter presents highly personalized cognitive products.

\subsection{Theoretical Perspectives: Activity Theory and Distributed Cognition}

We introduce Activity Theory (AT) and Distributed Cognition (DCog), two highly influential theoretical frameworks in the field of human-computer interaction, to provide a deeper interpretation of the design and contribution of \texttt{Irec}.

\subsubsection{Activity Theory (AT)}
Activity Theory provides powerful conceptual tools for analyzing complex human activities. It proposes a hierarchical structure of activity: the top level, \textbf{Activity}, is driven by collective motives; the middle level, \textbf{Action}, is guided by individual goals; and the bottom level, \textbf{Operation}, is completed unconsciously under specific conditions.

In the context of \texttt{Irec}, we can deconstruct the learning process as follows: "mastering calculus" is the top-level "activity," with motives such as acquiring knowledge and passing exams; "solving a specific indefinite integral problem" is the middle-level "action," with the goal of finding the correct answer; and typing the problem, clicking buttons, reading insights, etc., are the bottom-level unconscious "operations." The "Insight Recall" paradigm acts precisely at the junction of the "action" and "activity" levels. By connecting the current isolated problem-solving "action" with past relevant problem-solving experiences, it helps learners to go beyond the goal of a single problem, to reflect on and generalize more universal principles, thus serving the grand motive of the top-level "activity"—to achieve true knowledge internalization and transfer.

Furthermore, Activity Theory emphasizes the analysis of "contradictions" within the activity system. The design of \texttt{Irec} can be seen as a solution to a core contradiction in the learning activity: when the learner (subject) pursues deep understanding (object), the functions of the traditional tools they use (mediating artifacts, such as standard textbooks or de-contextualized review software) cannot meet their needs, leading to a conflict. \texttt{Irec} as a new type of "mediating artifact" intervenes in this system to help resolve this contradiction by providing contextualized, personalized historical insights. Reframing the system's design motivation in the language of Activity Theory will provide it with a richer and more profound theoretical foundation.

\subsubsection{Distributed Cognition (DCog)}
Distributed Cognition theory argues that cognitive processes do not occur solely within the individual brain but are distributed across the entire system composed of people, tools, and the environment. It expands the unit of cognition from the "individual's brain" to the "human-tool system."

From the perspective of Distributed Cognition, \texttt{Irec}'s knowledge graph is no longer just a database, but an "external cognitive artifact" of the learner's cognitive system. It materializes the learner's past thought processes and "aha moments." The process of "Insight Recall" is essentially a process of coordinating the flow of information between the learner's \textbf{internal cognition} (current thoughts, encountered confusion) and this \textbf{external cognitive resource} (past thoughts structurally stored in the graph).

This perspective elevates the \texttt{Irec} system from an auxiliary "scaffold" to a "cognitive partner" with humans. The system no longer unilaterally "supports" the user but participates as an indispensable part of the user's cognitive system in problem-solving and meaning-making. This theoretical positioning is highly consistent with the trend of "human-computer collaboration" in contemporary human-computer interaction research and can greatly enhance the theoretical foresight of the manuscript.

\subsection{Formal Modeling: A JITAI Model for Learning}
To translate the concept of "Insight Recall" into an operational intervention, we adopt and extend the \textbf{Just-in-Time Adaptive Intervention (JITAI)} model, which originated in the mobile health domain. JITAI is a powerful intervention design framework aimed at "providing the right support at the right time when it is most needed and most likely to be received, by adapting to the individual's changing state and context." Its dynamic, context-aware nature aligns perfectly with the philosophy of "Insight Recall."
The JITAI framework requires us to explicitly define the six core components of an intervention. The table below shows the specific mapping of the \texttt{Irec} system to the JITAI framework.

\begin{center}
\begin{longtable}{|p{0.2\textwidth}|p{0.3\textwidth}|p{0.45\textwidth}|}
\caption{JITAI Model of the Irec System} \label{tab:jitai_en} \\
\hline
\textbf{JITAI Component} & \textbf{Irec System Implementation} & \textbf{Description and Rationale} \\
\hline
\endhead

\textbf{Distal Outcome} & To enhance the networked integration of knowledge and the transferability of problem-solving skills. & The long-term educational goal of the system is to help learners build a stable, interconnected knowledge structure, enabling them to transfer skills to new contexts. This is one of the ultimate goals of self-regulated learning. \\
\hline
\textbf{Proximal Outcome} & To trigger metacognitive reflection (e.g., comparing, abstracting, self-explaining). & The immediate cognitive effect the intervention aims to achieve. The goal is to make the learner pause and think about their own thinking process, which is central to SRL. A successful "recall" will prompt generative learning activities. \\
\hline
\textbf{Decision Point} & The user submits a new problem or query to the system. & The specific moment when the system must decide whether to intervene. This is a clear, user-event-driven trigger. A future research direction is to explore more proactive decision points based on user state inference. \\
\hline
\textbf{Intervention Options} & 1. Present relevant insights: Display one or more historical \texttt{ProblemCard} nodes. \newline 2. Provide nothing: If the decision rule's confidence is below a threshold. & The set of discrete actions the system can take. "Providing nothing" is a crucial part of JITAI, to avoid unnecessary cognitive disruption when there is no high-quality match. \\
\hline
\textbf{Tailoring Variables} & - The text content of the new problem and its semantic embedding vector. \newline - The complete state of the user's knowledge graph. \newline - The user's selected learning mode and filtering level. & Real-time contextual data used to personalize the intervention. Currently, these are mainly task-related variables. Future research should explore incorporating user state variables like cognitive load and emotion. \\
\hline
\textbf{Decision Rule} & A personalized decision rule composed of a three-path hybrid recall mechanism, a multi-signal reranking model dynamically adjustable by learning mode, and an LLM similarity assessment module. & The core algorithm that links tailoring variables to intervention options. It is the "brain" of JITAI, responsible for calculating which intervention option is most likely to achieve the proximal outcome in the current context. \\
\hline
\end{longtable}
\end{center}

\section{\texttt{Irec} System Architecture and Implementation}
To transform the JITAI model of "Insight Recall" from a theoretical concept into an operational, extensible, and robust software system, we designed and implemented the \texttt{Irec} prototype system. This chapter will detail \texttt{Irec}'s system architecture, core workflow, key technical implementation details, and the engineering considerations made to ensure system robustness, thereby demonstrating how the theoretical framework is effectively operationalized.

\subsection{Overall Architecture}
\texttt{Irec} adopts a modular, layered software architecture aimed at achieving high cohesion, loose coupling, and clear separation of responsibilities. The architecture is logically divided into a Data Layer, Core Service Layer, Business Logic Layer, and Interface Layer.

\begin{itemize}[leftmargin=*]
    \item \textbf{Data Layer:} As the knowledge foundation of the system, the data layer uses a Neo4j graph database for persistent storage. The choice of a graph database is a core decision, as it can model and query the highly connected data relied upon by "Insight Recall" with extremely high efficiency. During system initialization, a series of database constraints and indexes are automatically established, including ID uniqueness constraints for \texttt{ProblemCard} and \texttt{Tag} nodes, as well as specialized Full-text and Vector Indexes for content retrieval and tag searching. This "Schema-First" design ensures data consistency and query performance from the ground up.

    \item \textbf{Core \& Business Logic Layers:} The system services are carefully divided into two levels. The Core Service Layer encapsulates interactions with external systems and low-level data operations, for example:
    \begin{itemize}
        \item \texttt{ConnectionPoolManager}: To ensure system scalability and efficient management of database resources, we implemented a global singleton Neo4j connection pool. All requests to the database go through this pool, significantly reducing the overhead of establishing and tearing down connections, which is a cornerstone of the system's high performance.
        \item \texttt{GenericGraphService}: Provides a set of generic, type-safe CRUD (Create, Read, Update, Delete) interfaces, decoupling the upper business logic from specific Cypher queries, making the knowledge graph model easy to extend.
        \item \texttt{SecureGraphConfig}: All sensitive system configurations (such as database passwords) are handled through an encryption service and stored in an encrypted format, reflecting adherence to security best practices.
    \end{itemize}
    The Business Logic Layer is built on top of the core services and implements the core paradigm of \texttt{Irec}. Its key services include the \texttt{SearchService}, responsible for the core search algorithm of "Insight Recall," the \texttt{IntelligentTagMapper} for automated tag mapping, and the \texttt{MathWorkflowService} which orchestrates the complete user interaction flow.

    \item \textbf{Interface Layer:} As the gateway for interaction between the system and the outside world, the interface layer currently consists of a series of command handlers based on the Tauri framework. This design allows \texttt{Irec}'s backend core capabilities to be seamlessly integrated into modern, cross-platform desktop applications, providing users with a smooth, low-latency interactive experience.
\end{itemize}

\subsection{Core Workflow: Parallelized Orchestration from Context to Reflection}
Unlike the traditional "request-response" model, the core interaction of \texttt{Irec} is designed as a structured workflow orchestrated by the \texttt{MathWorkflowService}, supporting parallel processing. The primary goal of this design is to minimize user-perceived latency. When a user inputs a new problem, the system initiates a session containing multiple parallel tasks, ensuring that both quick feedback and deep computation can proceed simultaneously.

A typical parallel workflow is as follows:
\begin{enumerate}
    \item \textbf{Context Input and Parallel Analysis:} User input (whether manually typed or from OCR) immediately triggers two parallel core tasks: an LLM call for deep semantic analysis of the input content, and a knowledge graph search based on initial keywords.
    \item \textbf{Instant Feedback and Background Deepening:} Preliminary, low-latency search results (such as full-text matches) can be immediately presented to the user to meet their immediate needs. Meanwhile, more computationally expensive tasks, such as vector generation, deep semantic analysis, and AI recommendations, are executed asynchronously in the background.
    \item \textbf{Progressive Enhancement and Synthesis of Results:} As background tasks (such as intelligent tag mapping, AI recommendation analysis) complete, their results are progressively pushed to the front-end interface via an event stream. For example, the system first displays a preliminary list of similar problems, then updates these problems with precise tags, and finally presents the deep recommendation relationships generated by AI.
\end{enumerate}

\subsection{Key Technologies and Algorithm Details}
The core competitiveness of the \texttt{Irec} system stems from its series of collaborating algorithms and technical implementations.

\subsubsection{Context-Aware Retrieval: Three-Path Hybrid Recall and Fusion}
To maximally simulate the complex associative networks of human thought, the system's \texttt{SearchService} implements an innovative \textbf{Three-Path Hybrid Recall} architecture. This architecture executes three independent recall channels in parallel:
\begin{enumerate}
    \item \textbf{Vector Similarity Recall:} By converting the user query into a high-dimensional semantic vector, it performs an efficient approximate nearest neighbor search within Neo4j's built-in vector index to capture deep semantic relevance.
    \item \textbf{Full-text Keyword Recall:} Utilizes the database's native full-text index for precise lexical matching of problem and insight content to capture surface-level lexical relevance.
    \item \textbf{Tag Hierarchy Expansion Recall:} This path first identifies one or more "entry tags" in the tag knowledge system based on the query text, then traverses the graph along the \texttt{PARENT\_OF} relationship to collect all relevant child tags, and finally retrieves all problem cards associated with this expanded tag set as candidates. This captures the structural relevance of knowledge.
\end{enumerate}
After the recall phase, the system enters a \textbf{Fusion} stage. Results from the three channels are first mapped to a comparable range through their respective score normalization algorithms (e.g., using Z-Score normalization for vector similarity, and Min-Max normalization for full-text search scores). Then, the system merges the normalized scores with predefined weights (\texttt{MergeWeights}) and gives a \textbf{Multi-match Bonus} to items recalled by multiple channels, producing a unified, preliminarily ranked candidate set.

\subsubsection{Personalized Learning Modes and Multi-Signal Reranking}
Based on the preliminarily fused candidate set, the system enters a more refined \textbf{Multi-Signal Reranking} stage. The goal of this stage is to go beyond simple content relevance and respond deeply to the user's immediate learning intent. The final ranking score $S_{\text{final}}(c_i)$ for each problem card $c_i$ in the candidate set is defined by the following general model:
\begin{equation}
S_{\text{final}}(c_i) = w_{\text{rel}} \cdot R(c_i) + w_{\text{acc}} \cdot A(c_i, M) + w_{\text{temp}} \cdot T(c_i, M) + w_{\text{div}} \cdot D(c_i)
\end{equation}
where:
\begin{itemize}[nosep, leftmargin=*]
    \item $c_i$ represents the $i$-th problem card in the candidate set.
    \item $w_{\text{rel}}, w_{\text{acc}}, w_{\text{temp}}, w_{\text{div}}$ are the weight coefficients for content relevance, access frequency, temporal recency, and diversity signals, respectively.
    \item $R(c_i)$ is the content relevance score, which is the base score after three-path recall fusion and normalization.
    \item $A(c_i, M)$ is the access frequency score, calculated based on the learning mode $M$.
    \item $T(c_i, M)$ is the temporal recency score, also calculated based on the learning mode $M$.
    \item $D(c_i)$ is the diversity bonus score.
    \item $M$ represents the user-selected personalized learning mode (\texttt{Learning}, \texttt{Review}, \texttt{Balanced}).
\end{itemize}

\textbf{Mathematical Implementation of Personalized Learning Modes:}

The system's personalization capabilities are mainly achieved by dynamically adjusting the calculation of signal functions $A(c_i, M)$, $T(c_i, M)$ and the weights of each signal according to the learning mode $M$.

\textbf{Access Frequency Score $A(c_i, M)$}

Let $N_{\text{access}}(c_i)$ be the historical access count of card $c_i$.
\begin{itemize}
    \item When the learning mode $M$ is \textbf{Learning} or \textbf{Balanced}, the system prefers to recommend popular content. The access frequency score is modeled as a logarithmic growth function to smooth the impact of high access counts:
    \[ A(c_i, M_{\text{learn}}) = \frac{\ln(1 + N_{\text{access}}(c_i))}{\ln(1 + K_{\text{acc}})} \]
    where $K_{\text{acc}}$ is a scaling constant for access count normalization (10.0 in the implementation).

    \item When the learning mode $M$ is \textbf{Review}, the system aims to reinforce weak areas and prefers to recommend less frequently accessed content. The access frequency score is modeled as a reverse-decreasing function:
    \[ A(c_i, M_{\text{review}}) = \frac{1}{1 + N_{\text{access}}(c_i) / K_{\text{acc}}} \]
\end{itemize}

\textbf{Temporal Recency Score $T(c_i, M)$}

Let $\Delta t(c_i)$ be the number of days since the last access to card $c_i$.
\begin{itemize}
    \item When the learning mode $M$ is \textbf{Learning} or \textbf{Balanced}, the system prefers to recommend recent content. The temporal recency score is modeled as a decay function:
    \[ T(c_i, M_{\text{learn}}) = \frac{1}{1 + \Delta t(c_i) / T_{\text{half}}} \]
    where $T_{\text{half}}$ is a time half-life constant (30.0 in the implementation) that controls the rate of score decay over time.

    \item When the learning mode $M$ is \textbf{Review}, the system aims to resurface forgotten content, with logic similar to a forgetting curve. Thus, the temporal recency score is modeled as an S-shaped growth function; the older the item, the higher the score, representing a stronger need for review:
    \[ T(c_i, M_{\text{review}}) = \frac{\Delta t(c_i) / T_{\text{half}}}{1 + \Delta t(c_i) / T_{\text{half}}} \]
\end{itemize}

\textbf{Diversity Bonus Score $D(c_i)$}

Let $|P(c_i)|$ be the size of the set of recall paths for card $c_i$ (e.g., \{\texttt{vector}, \texttt{fulltext}\}). The diversity bonus aims to reward results found by multiple independent recall channels.
\[ D(c_i) = \frac{|P(c_i)| - 1}{N_{\text{paths}} - 1} \]
where $|P(c_i)|$ is the number of recall paths, and $N_{\text{paths}}$ is the total number of recall channels (3 in this system).

\textbf{Weight Configuration Table:}
\begin{center}
\begin{tabular}{|l|c|c|c|c|}
\hline
\textbf{Learning Mode (M)} & \textbf{$w_{\text{rel}}$} & \textbf{$w_{\text{acc}}$} & \textbf{$w_{\text{temp}}$} & \textbf{$w_{\text{div}}$} \\
\hline
Learning & 0.50 & 0.20 & 0.20 & 0.10 \\
\hline
Review & 0.40 & 0.25 & 0.25 & 0.10 \\
\hline
Balanced & 0.60 & 0.15 & 0.15 & 0.10 \\
\hline
\end{tabular}
\end{center}

\subsubsection{Intelligent Tag Mapping}
To address the core pain point of "high manual maintenance costs" in personal knowledge management tools, \texttt{Irec} implements a highly automated \texttt{IntelligentTagMapper} service. This service is not defined by a single closed-form formula but is a multi-stage \textbf{Heuristic Decision Algorithm}.

It employs an innovative \textbf{Two-Phase, Context-focused Batch Processing} pipeline to efficiently and accurately map loose, AI-suggested tags to the existing hierarchical structure in the knowledge graph:
\begin{enumerate}
    \item \textbf{Phase 1: Batch Branch Selection.} The system integrates a batch of pending tag suggestions with the problem context and, through a single LLM call, selects one or more of the most relevant high-level knowledge branches for each tag (e.g., "Calculus" or "Linear Algebra").
    \item \textbf{Phase 2: Batch Precise Selection.} After determining the macro branches, the system packages each tag with its corresponding branch context again and, through a second LLM call, makes a more fine-grained decision: whether to map it to an existing tag within the branch or to create a new tag under a parent node in that branch.
\end{enumerate}

Although the core of this process is driven by a large language model, its key stages, such as candidate tag pre-screening and decision fallback, follow explicit scoring models.
\begin{itemize}
    \item \textbf{Candidate Tag Pre-screening:} For a user-suggested tag $T_{\text{new}}$, the system first screens the existing tag library to find a Top-N candidate set $C$. The initial confidence score $Score_{\text{cand}}(T_j)$ for a candidate tag $T_j \in C$ is determined by its name similarity to the new tag and its semantic relevance to the problem context.
    \[ Score_{\text{cand}}(T_j) = w_{\text{name}} \cdot \text{Sim}_{\text{cosine}}(E_{T_{\text{new}}}, E_{T_j}) + w_{\text{context}} \cdot \text{Sim}_{\text{cosine}}(E_{\text{problem}}, E_{T_j}) \]
    where:
    \begin{itemize}[nosep, leftmargin=2em]
        \item $E_{T_{\text{new}}}, E_{T_j}, E_{\text{problem}}$ are the semantic embedding vectors for the new tag, candidate tag, and problem context, respectively.
        \item $\text{Sim}_{\text{cosine}}$ is the cosine similarity function.
        \item $w_{\text{name}}$ and $w_{\text{context}}$ are weight coefficients balancing the importance of tag semantics and problem context (0.7 and 0.3 in the implementation).
    \end{itemize}

    \item \textbf{Intelligent Fallback Strategy:} In cases where the LLM call fails or returns low-quality results, the system activates an \textbf{Intelligent Fallback Strategy}. This strategy selects the optimal candidate from the candidate set using the following scoring function:
    \[ Score_{\text{fallback}}(T_j) = w_{\text{conf}} \cdot \text{Conf}(T_j) + w_{\text{level}} \cdot W_{\text{level}}(L(T_j)) \]
    where:
    \begin{itemize}[nosep, leftmargin=2em]
        \item $\text{Conf}(T_j)$ is the confidence score $Score_{\text{cand}}(T_j)$ obtained by the candidate tag $T_j$ during pre-screening.
        \item $L(T_j)$ is the depth (level) of the candidate tag $T_j$ in the knowledge hierarchy.
        \item $W_{\text{level}}$ is a weight function that is negatively correlated with hierarchy depth, aiming to prioritize higher-level (more stable) tags as fallback options. For example, it can be defined as $W_{\text{level}}(l) = 1 / (1+l)$.
        \item $w_{\text{conf}}$ and $w_{\text{level}}$ are the corresponding weight coefficients (0.7 and 0.3 in the implementation).
    \end{itemize}
\end{itemize}

\subsection{Engineering Robustness and Performance Considerations}
In addition to the innovation of its core algorithms, the implementation of the \texttt{Irec} system includes a large number of engineering practices aimed at ensuring its stable and efficient operation in real-world usage scenarios.
\begin{itemize}[leftmargin=*]
    \item \textbf{Comprehensive Asynchronous Processing:} The core operations of the system, especially those involving network I/O (like database queries) or high-latency computations (like LLM calls), use an asynchronous execution model. For example, the vector embedding generation for problem cards is designed to be completed in a background task (\texttt{tokio::spawn}), ensuring the immediate responsiveness of the front-end interface, which does not freeze while waiting for AI calculations.
    \item \textbf{Systematic Logging and Performance Monitoring:} The system integrates a custom workflow logger (\texttt{WorkflowLogger}). It creates a detailed log with a unique session ID for each complete user interaction, accurately recording the execution time of each step, sub-operation, and even database query. This mechanism provides invaluable data for performance bottleneck analysis, debugging, and continuous optimization of the system.
    \item \textbf{Design and Testing for Scalability:} To verify the system's ability to handle large-scale data, we specifically designed and implemented a \texttt{BulkImportService}. This service supports the parallelized, batch import of thousands of problem nodes and provides real-time feedback on import progress, demonstrating that the system architecture fully considered future scalability from the outset.
    \item \textbf{Rigorous Automated Testing:} The system's codebase includes comprehensive unit and integration tests. In particular, the integration tests utilize the Testcontainers framework, which can dynamically launch a temporary, real Neo4j database container during test execution. This ensures that all database interaction logic is effectively validated in an environment close to production, guaranteeing the long-term stability and maintainability of the system.
\end{itemize}

\section{Illustrative Scenario: A Walkthrough of the Irec Workflow}
The purpose of this chapter is not to evaluate the actual impact of \texttt{Irec} on learning outcomes, but to illustrate, through a concrete, end-to-end scenario, how a user experiences a complete 'Insight Recall' process with the help of \texttt{Irec}, and to clarify the roles played by the system's various technical modules (such as hybrid retrieval, dynamic reranking) in this process.

\textbf{Scenario Background:} Alex is a university student studying calculus. He has previously used \texttt{Irec} to record an insight about solving an indefinite integral problem using the "u-substitution method."

\textbf{Step 1: Insight Capture and Knowledge Graph Evolution}

A few weeks ago, Alex successfully solved an indefinite integral problem $\int x(x^2+1)^3 dx$. He recorded his solution process and reflection in \texttt{Irec}: "The key to this problem is observing that one part of the integrand (x) is a multiple of the derivative of another part ($x^2+1$). This suggests I should use the \textbf{u-substitution method}, letting $u=x^2+1$."

\texttt{Irec}'s \texttt{InsightCaptureService} processes this input in the background. First, an LLM is called to parse this note into a structured JSON object containing the problem, insight, and suggested tags. Subsequently, the system creates a new \texttt{ProblemCard} node in the Neo4j database, and the \texttt{IntelligentTagMapper} service maps the tags to the graph, ultimately establishing a \texttt{HAS\_TAG} relationship between this new \texttt{ProblemCard} and the relevant \texttt{Tag} nodes.

\textbf{Step 2: Emergence of a New Context and the JITAI Decision Point}

A few weeks later, while studying definite integrals, Alex encounters a new difficult problem: calculating $\int_{0}^{\sqrt{\pi}} x\sin(x^2)dx$. He feels stuck because he is unfamiliar with this function form. He inputs this problem into \texttt{Irec}, seeking relevant historical experience. When inputting the problem, Alex chooses the default \textbf{Balanced Mode} and the more focused \textbf{Strict Mode} for filtering, as he wants to concentrate on solving the current type of problem.

Alex's input constitutes a \textbf{Decision Point} in the JITAI model. The system must immediately decide whether and how to intervene based on the current context.

\textbf{Step 3: Execution of the Decision Rule and Insight Recall}

\texttt{Irec}'s \texttt{SearchService} (the \textbf{Decision Rule}) is activated, using Alex's new problem and his chosen mode as \textbf{Tailoring Variables}. First, the \textbf{Three-Path Hybrid Recall} mechanism is activated, and the candidate set is personalized reranked using the weights for the "Balanced Mode", making the previous \texttt{ProblemCard} about u-substitution rank first. Before returning the results, the system enters the \textbf{AI Relationship Analysis} phase, sending the new problem and the top-ranked historical card to an LLM for analysis. The LLM's analysis returns a \textbf{Similarity Score} of 1 (Slight Variation). Subsequently, the system applies Alex's chosen \textbf{Strict Mode}, and since the score of 1 is less than or equal to the threshold of 2, this card passes the filter. Meanwhile, another card about double integrals is rated with a similarity of 3 and is therefore successfully filtered out.

The system immediately presents this filtered historical card as an \textbf{Intervention Option} on Alex's screen, titled "Related Insight: Regarding the U-Substitution Method." This is a complete "Insight Recall."

\textbf{Step 4: Metacognitive Scaffold Triggers Reflection (Proximal Outcome)}

Seeing his own "insight" from a few weeks ago, Alex's thinking is activated. This "recalled" insight now acts as an effective \textbf{Metacognitive Scaffold}. Alex's internal monologue is as follows:

"Oh, I remember this! The key was 'one part is the derivative of another part.' Let me look at the current problem... The derivative of $x^2$ is $2x$, and there's an $x$ right outside. The structure is exactly the same! Although one is a polynomial and the other is a trigonometric function, the core idea is universal. I should be able to use u-substitution here too, letting $u=x^2$."

In this process, Alex was not given the answer. Instead, he was guided to \textbf{Compare}, \textbf{Abstract}, and \textbf{Self-explain}. This active, generative reflection process, triggered by "Insight Recall," is precisely the \textbf{Proximal Outcome} we aim to achieve.

\textbf{Step 5: Knowledge Integration and Transfer (Distal Outcome)}

Using this reactivated idea, Alex successfully solves the new definite integral problem. More importantly, this experience elevates his understanding of the "u-substitution method" beyond specific function forms. He adds a new insight to the new problem card in \texttt{Irec}: "This experience taught me that the essence of the u-substitution method is to recognize the inverse 'chain rule' structure in a composite function, regardless of the specific function form it appears in."

Through this reflection facilitated by \texttt{Irec}, Alex not only solved a new problem but also connected two isolated problem-solving experiences, forming a more generalizable and transferable knowledge structure. This is precisely the \textbf{Distal Outcome} the system aims to achieve—promoting the networked integration and transfer of knowledge.

\textbf{(Extended Part) Scenario Extension: Deepening Understanding with the Guided Inquiry Module}

Suppose that in Step 4, despite seeing the historical insight, Alex is still unsure how to apply it to the new problem involving a trigonometric function. He can then click the "Explore with AI Tutor" button next to the insight card.

\texttt{Irec}'s \texttt{GuidedInquiryService} is activated, packaging Alex's current problem "calculate $\int x\sin(x^2) dx$" and the historical insight about "u-substitution" and sending them to an LLM specializing in mathematics.

A dialog box appears, and the AI tutor begins to ask: "Hi Alex, I see you successfully used u-substitution on a polynomial problem before, that's great! You summarized the key was 'one part is the derivative of another.' Can you find parts with a similar relationship in this current $\sin(x^2)$ problem?"

Through a few rounds of such dialogue, the AI tutor guides Alex to realize on his own the relationship between the derivative of $x^2$ and the other part of the function, $x$, thus independently transferring old knowledge to a new context. This conversation not only helps him solve the problem but also validates and reinforces his understanding of the core idea of the "u-substitution method," effectively overcoming the cognitive barrier caused by the different function forms.

This complete closed-loop and extended scenario demonstrates how the "Insight Recall" paradigm, through a JITAI model and an optional guided inquiry module, can seamlessly support and enhance a learner's self-regulated learning in a real learning workflow.

\section{Discussion: Limitations and Future Research Directions}
This paper has proposed and implemented a novel learning support paradigm—"Insight Recall"—and formalized it as a metacognitive scaffolding system based on the JITAI framework. Although the implementation of the \texttt{Irec} prototype system and the illustrative scenario have preliminarily demonstrated the feasibility of this paradigm, there are still some limitations worth discussing, which also point to directions for future research.

\subsection{Research Limitations}
First, the most significant limitation of this study is the \textbf{lack of large-scale, controlled empirical evaluation}. Although the illustrative scenario demonstrates the feasibility of \texttt{Irec}, the actual impact of the "Insight Recall" paradigm on learners' knowledge transfer ability and SRL skill improvement still needs to be tested through rigorous scientific experiments. For any academic contribution in the HCI or AIED fields, the lack of rigorous user studies is a major obstacle to publication. Therefore, this study is positioned as a conceptual and technical validation, aiming to lay the groundwork for future empirical research.

Second, the \textbf{dependency on large language models}. \texttt{Irec}'s automated knowledge graph construction process relies on the semantic understanding capabilities of LLMs. Although we have designed a \textbf{human-in-the-loop validation and correction mechanism} to mitigate this risk, the inaccuracy of LLM outputs remains a challenge for the system.

Third, the \textbf{limitation of cognitive fixation and depth of reflection}. The historical insights presented by the system may lead learners to fall into a mental rut or to reflect inefficiently or even incorrectly on an insight that was flawed in the first place. The "Guided Inquiry" module is a response to this problem, but its effectiveness also depends on the quality of interaction with the LLM.

Fourth, the \textbf{preliminary nature of the JITAI model implementation}. We mapped \texttt{Irec} to the JITAI framework, providing it with a rigorous theoretical model. However, the "decision rule" in the current prototype system is more "just-in-time" than "dynamically adaptive." The system's personalization is mainly reflected in its response to user-preset modes, rather than dynamic adaptation to the user's real-time state. This is a gap compared to the forefront of JITAI research.

Finally, from a technical implementation perspective, the \textbf{static nature of the reranking model} and the \textbf{relative simplicity of the retrieval model} are also areas for improvement. The current multi-signal reranking model relies on a set of static, preset weights; future systems should be able to dynamically learn these weights. Also, while the three-path hybrid recall mechanism is effective, there is still room for improvement compared to the state-of-the-art models in the information retrieval field.

\subsection{Cognitive Load Considerations}
Although \texttt{Irec} aims to reduce extraneous cognitive load by automating knowledge graph construction, its core interaction design, such as human-in-the-loop validation and guided inquiry, inevitably requires users to invest mental effort. This investment is not a design flaw of the system, but a crucial trade-off. The design philosophy of \texttt{Irec} is not to blindly pursue "effortlessness" or "full automation," but to seek a dynamic balance between "automation convenience" and "beneficial cognitive investment."

The human-in-the-loop validation process requires the user to adjudicate AI suggestions, and this process itself may constitute a germane cognitive load, as it prompts the user to think precisely about the classification and association of knowledge, thereby deepening understanding. Similarly, the "Guided Inquiry" module actively increases the depth of cognitive processing through Socratic dialogue, with the precise aim of stimulating higher-order thinking activities. How to design interfaces and interactions to minimize unnecessary extraneous load while maximizing the effectiveness of germane load is a core issue in HCI and an important direction for future iterations of \texttt{Irec}.

\subsection{Roadmap for Future Research}
Based on the solid architecture and theoretical foundation of \texttt{Irec}, we propose a roadmap for future research, which is both a vision for future work and a call for collaboration with the research community.

First, \textbf{core empirical evaluation} is the highest priority. A longitudinal controlled experiment lasting several weeks could be conducted, dividing subjects into an \texttt{Irec} experimental group, a PKM control group, and an SRS control group. Combining quantitative (e.g., knowledge transfer tests) and qualitative (e.g., think-aloud protocols) data, we can systematically examine the effectiveness of the "Insight Recall" paradigm.

Second, future technical iterations can explore the following directions:
\begin{itemize}[leftmargin=*]
    \item \textbf{Moving towards truly adaptive interventions:} This is a core breakthrough from the preliminary implementation of the current JITAI model.
    \begin{itemize}
        \item \textit{From passive triggers to proactive interventions:} Future versions of \texttt{Irec} could explore more proactive intervention timings. For example, by passively sensing user interaction behaviors (such as pausing for a long time on a problem, scrolling back and forth, highlighting and then un-highlighting) to infer that they are in a state of being "stuck" or "confused," and then proactively providing relevant "Insight Recall" in this "window of opportunity," even if the user has not explicitly requested help.
        \item \textit{Integrating richer user state variables:} To make interventions truly "adaptive" to the "person" and not just the "problem," future systems could incorporate deeper user state variables. For example, using sensor-free affect detection by analyzing interaction logs, or even integrating real-time cognitive load monitoring data from brain-computer interfaces (BCI), to achieve ultimate personalization of intervention content and timing.
    \end{itemize}
    \item \textbf{Model self-optimization using user correction data:} Every correction a user makes to AI-generated content is a valuable feedback signal. Future research can explore how to use this data to continuously fine-tune the language model, allowing the system to gradually learn the user's personal expression habits and knowledge organization patterns.
    \item \textbf{Enhancing support for collaborative learning scenarios:} Future versions could be extended to support collaborative environments like teams or classrooms, allowing users to selectively share "insights" and recall relevant, anonymized insights from peers, adding a powerful social learning dimension to the paradigm.
    \item \textbf{Integrating richer learning analytics and visualizations:} Future \texttt{Irec} systems could integrate more powerful learning analytics modules to provide users with visual reports on their knowledge network structure, concept evolution paths, and cognitive blind spots (such as "structural holes" in the knowledge graph), helping users to better navigate and understand their "second brain."
\end{itemize}

\section{Conclusion}
This paper aims to address the challenges modern learners face in integrating complex knowledge networks and engaging in effective self-regulated learning. We have introduced and systematically elaborated on "Insight Recall," a novel learning support paradigm that conceptualizes the context-triggered retrieval of personal historical insights as an automated scaffold for promoting metacognitive reflection. Through the innovative application of Activity Theory, Distributed Cognition, and the Just-in-Time Adaptive Intervention (JITAI) framework, we have provided a profound theoretical interpretation and a rigorous formal model for this paradigm, and have designed and implemented the prototype system \texttt{Irec} to operationalize it. \texttt{Irec}, through human-AI collaborative automated knowledge graph construction and a multi-strategy hybrid retrieval mechanism, demonstrates a viable path to achieving precise, timely metacognitive support while reducing the user's cognitive load.

We believe the contribution of this paper extends beyond the implementation of a single system. It provides the academic and practitioner communities with a new set of theoretical perspectives and a design language for thinking about and building the next generation of intelligent learning environments. Future learning tools should not merely be transmitters of knowledge or supervisors of memory, but should become partners in the learner's cognitive process, catalysts for reflection, and enablers of self-discovery. By placing the learner's own thinking at the core of the learning loop, the "Insight Recall" paradigm has taken a solid step towards realizing this vision.

\subsection*{An Invitation for Empirical Validation and Collaboration}
The conceptual framework and prototype system proposed in this paper lay the foundation for a new method of self-regulated learning. However, validating the true utility of the "Insight Recall" paradigm requires rigorous longitudinal studies on diverse learner populations, which is beyond the current resources of the author as an undergraduate researcher.

Therefore, this research is not a final conclusion, but a foundational blueprint and a robust platform intended to inspire future research. We hereby openly and sincerely invite researchers, laboratories, and teams with expertise in Human-Computer Interaction (HCI), AI in Education (AIED), and Learning Sciences to join in the challenge of empirically testing, critiquing, and extending the \texttt{Irec} system. The author is eager to participate in collaborations and contribute to the collective research effort aimed at understanding and enhancing technology-mediated learning.

\subsection*{Code Availability and Reproducibility}
To promote the reproducibility and extension of this research, we are preparing a comprehensive open-source release for the \texttt{Irec} prototype. The upcoming version (v2) of this paper will feature a public GitHub repository containing the full source code. 

To further facilitate replication and experimentation by the community, this release will be accompanied by a detailed reproducibility package. This package will include key implementation details, such as the specific prompts engineered for the LLM modules, recommendations for the language and embedding models tested, and a complete guide for setting up the system environment. We welcome the community to follow our progress and engage with the project.

\end{document}